\documentstyle[11pt,twoside,pasp3D,epsf]{article}

\begin{document}

\title{An Imaging Fourier Transform Spectrometer for 
the Next Generation Space Telescope}

\author{James. R. Graham}
\affil{Department of Astronomy, University of California, Berkeley, CA 94720}

\begin{abstract}

Due to its simultaneous deep imaging and integral field spectroscopic
capability, an Imaging Fourier Transform Spectrograph (IFTS) is
ideally suited to the Next Generation Space Telescope (NGST) mission,
and offers opportunities for tremendous scientific return in many
fields of astrophysical inquiry.  We describe the operation and
quantify the advantages of an IFTS for space applications. The
conceptual design of the Integral Field Infrared Spectrograph (IFIRS)
is a wide field ($5'.3 \times 5'.3$) four-port imaging Michelson
interferometer.
\end{abstract}

\section{Introduction}
When the Next Generation Space Telescope (NGST) begins observing
towards the end of the first decade of the next millennium it will
usher in a new era of infrared (IR) astronomy.  The combination of a
deployable 8-meter aperture and an L2 orbit, which will allow the
telescope to cool to 30-50K, will enable zodiacal-light limited
performance for $\lambda < 10~\mu$m.

The rationale for placing NGST far from Earth and shielded from the
Sun is reduction of the background by factors of up to 10$^6$ compared
to terrestrial environments. This will enable studies of the origins of
the structure in the universe, galaxies and quasars, stars, and
planetary disks --- core mission objectives of the NASA Origins
Program. Many of these programs require IR observations:  the light
from distant galaxies seen at early epochs is redshifted by the
expansion of the universe; regions of star formation in our galaxy are
obscured by dust in the visible, but are penetrated by infrared
radiation; and cool objects such as forming stars and proto-planetary
disks emit in the IR.

\section{What is IFIRS?}
\begin{figure}
\plottwo{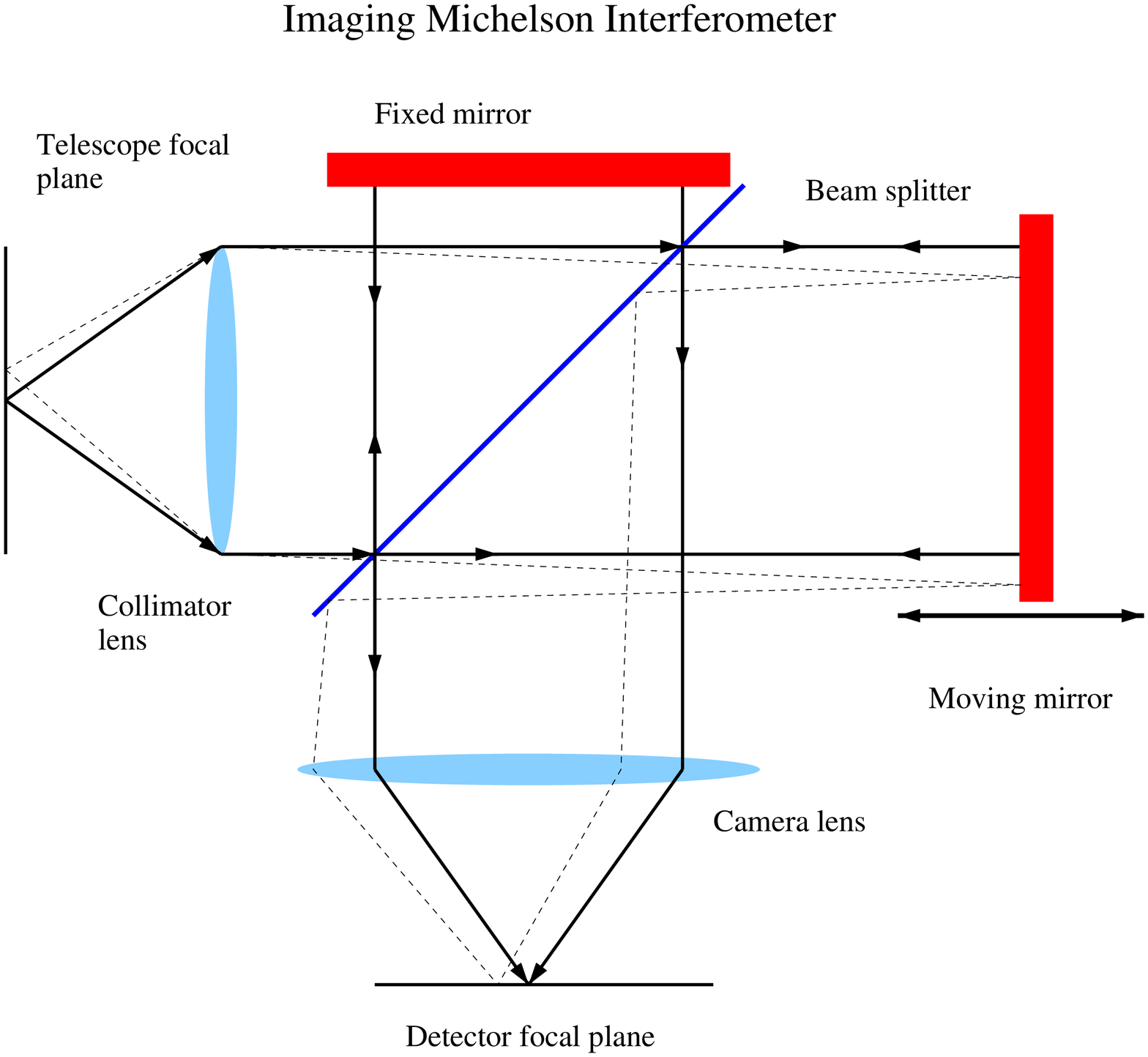}{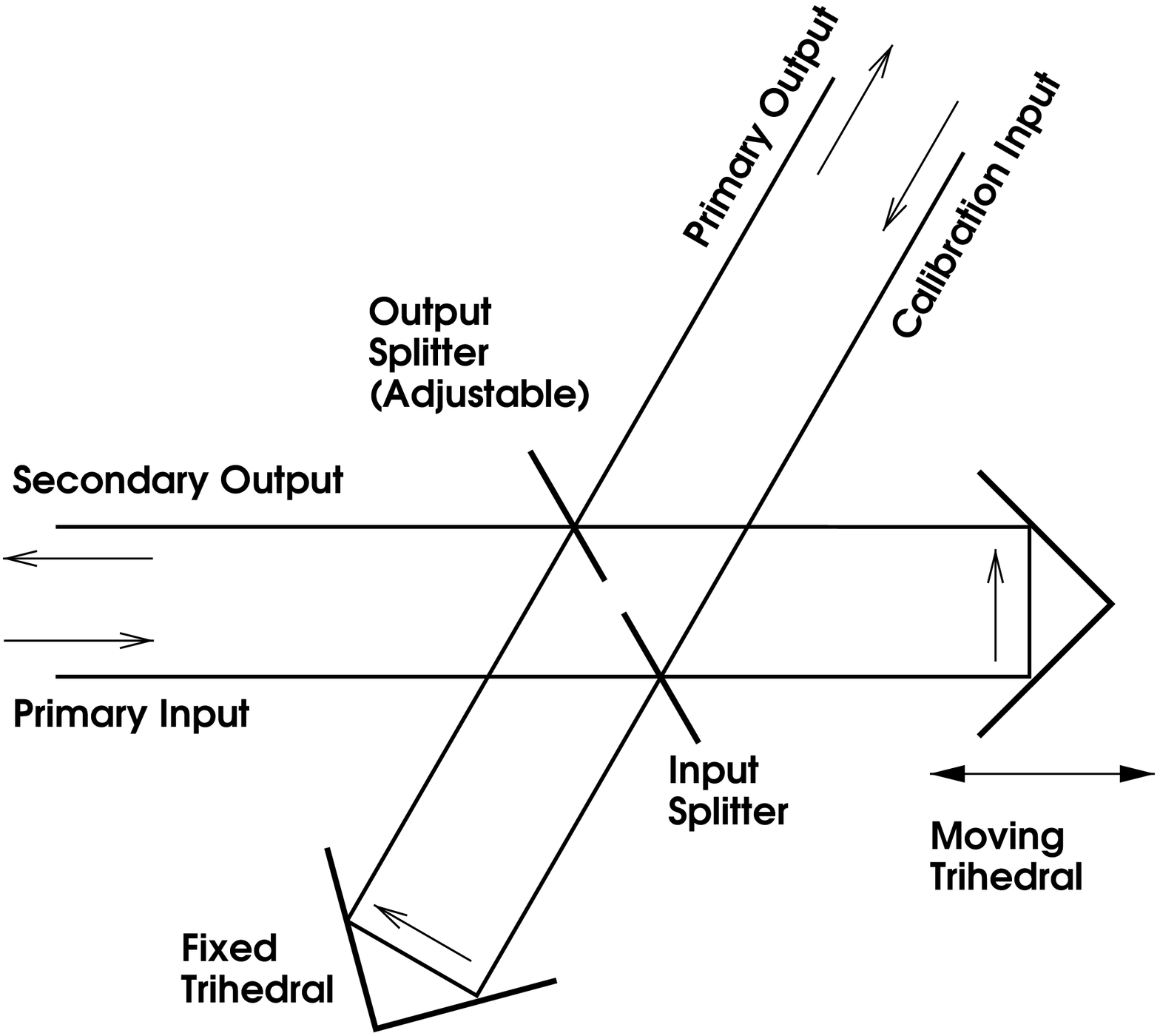}
\caption{{\bf Left:} An IFTS is a Michelson interferometer where the
telescope focal plane is imaged onto a detector array. An
interferogram is recorded for every pixel, and hence a spectrum can be
obtained for every object. {\bf Right:} IFIRS has a four-port design
that wastes none of the light. Instead of flat mirrors, IFIRS uses
cube-corners, which displace the input
and output beams.  }
\label{imaging.michelson}
\end{figure}

The goal of IFIRS is to obtain ultra-deep, wide field, diffraction
limited imagery from near- to mid-IR wavelengths, with flexible
spectral resolution (Graham et al. 1998).  IFIRS is a Michelson
interferometer configured as an imaging Fourier transform spectrometer
(IFTS) (See Fig.  \ref{imaging.michelson}) (Bennett 1993).  An
interferogram is recorded for every pixel in the field of view, and
hence a spectrum can be obtained for every object.  Since IFIRS is an
FTS, it is both a high-throughput camera and a multi-object or
integral field spectrometer. As a camera IFIRS has unusual flexibility
of spectral resolution, and a unique pan-chromatic imaging mode.
IFIRS is a four-port interferometer (Fig. \ref{imaging.michelson}).
One of the disadvantages of a classical flat-mirror Michelson is that
the beam-splitter reflects 50\% of the light back to the source; a
four-port design collects all the light.

\subsection{Design of IFIRS}

On the object side a collimator illuminates the interferometer with
parallel light. The interfering beams are collected by a camera,
creating a one-to-one mapping between points in the object and image
planes.  By placing a detector focal plane array (FPA) at the focus of
the camera, each pixel is matched to a single point on the sky.  At
any given optical path difference (OPD) the image of the sky is
modulated spatially by the interferometer's fringe pattern, which
encodes the spectral information. By recording images of the sky at
different OPDs, the spectrum of each pixel can be reconstructed. The
OPD is scanned in discrete steps since FPAs are integrating detectors.
The time series from each pixel forms an independent interferogram for
every point on the sky within the field of view.  These interferograms
are Fourier transformed individually yielding a spectral data cube
composed of the same spatial elements as the image. The sampling
theorem establishes the number and amplitude of OPD steps necessary to
recover the spectrum at a given spectral resolution.

\begin{figure}
\plotfiddle{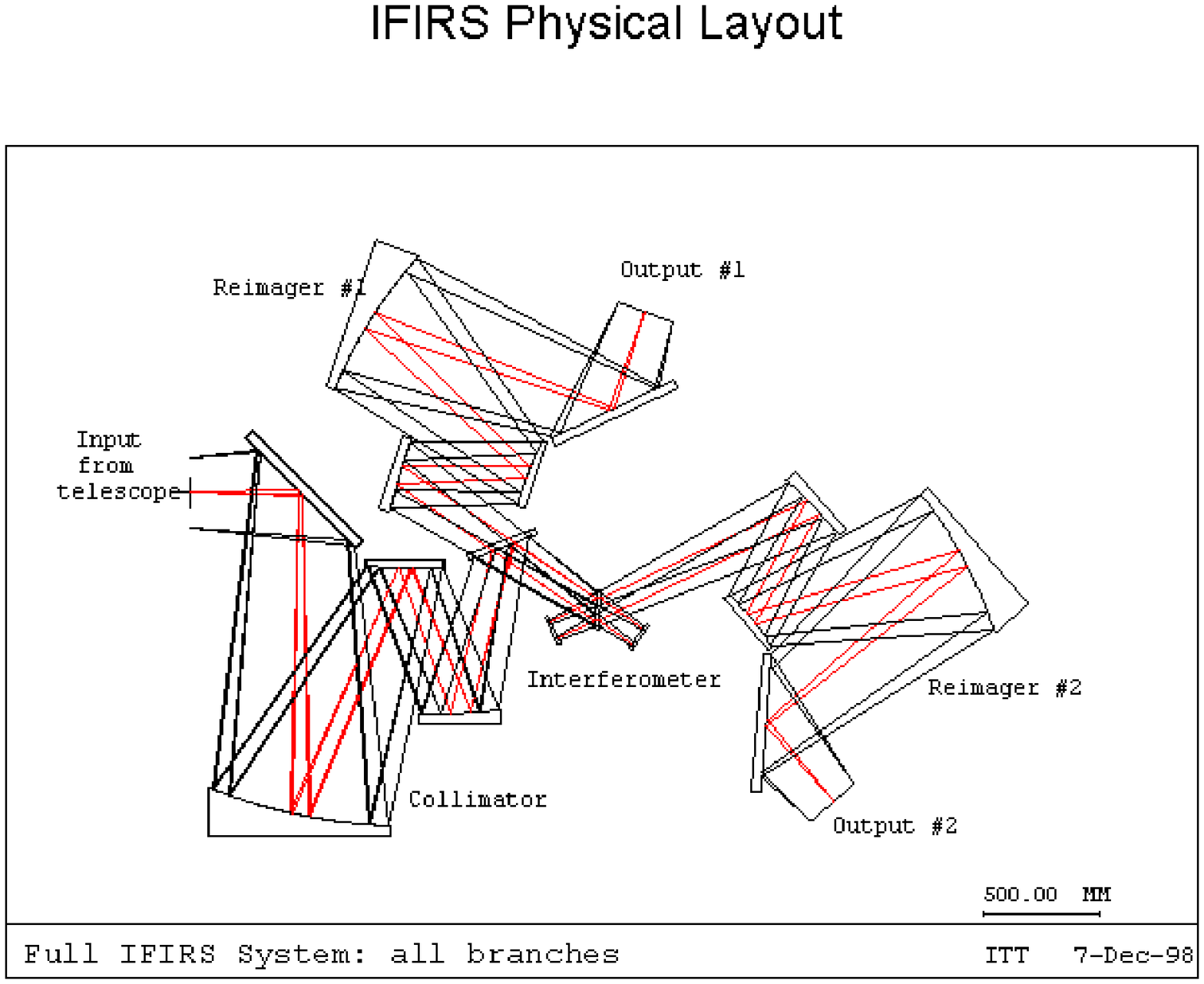}{2.3in}{0}{50}{50}{-160}{0}   
\caption{The optical layout of IFIRS.  For simplicity only the near-IR
FPA have been shown at the output ports. The final fold mirror is
either a dichroic or flip-mirror which directs the mid-IR light to the
mid-IR detectors. For clarity reimager \# 2 has been rotated by 180
degrees.  The actual design is symmetric about the beam-splitter.}
\label{wiregrid1}
\end{figure}

The major components of IFIRS, shown in Fig. \ref{wiregrid1}, are the
collimator, interferometer, cameras, and FPAs and associated
electronics.  Two important ancillary subsystems, not shown are the
metrology system and the calibration unit.

The collimator is a three mirror anastigmat (TMA) that illuminates a
four-port Michelson. There are two input and two output ports. One
input port is fed with the sky signal, and the other input can be
illuminated by the calibration unit. The interferometer consists of a
50:50 reflecting/transmitting beam splitter, and two cube-corner
retroreflectors.  The appropriate beam splitter is selected using a
filter-wheel mechanism. One cube-corner is located on a translation
stage, which permits precision control of the OPD. The OPD is
monitored using an interferometric metrology system. Each output port
of the interferometer feeds identical TMA cameras. A final reflection
(an articulated fold mirror or a static dichroic) directs short
wavelength radiation to the near-IR FPA and long wavelength radiation
to the mid-IR FPA. In summary, the properties of IFIRS are:

The absolute wavelength calibration is provided by the interferometer
metrology system, which is referenced to a diode laser.  The accuracy
should be better than 7.5 GHz (0.25 wavenumbers).  This provides the
wavelength calibration for both pure and dispersed FTS modes.

{\small
\begin{center}
\begin{table}
\caption{Observational Capabilities}
\medskip
\begin{tabular}{lll}
\hline
		& NIR Channel	& MIR Channel \\
\hline
Bandpass/Detector	& 0.6-5.6 $\mu$m/InSb & 5-15 $\mu$m/HgCdTe \\
Maximum Spectral Resolution 	& 1 cm$^{-1}$	& 1 cm$^{-1}$ \\ 
FOV/Array Format	& $5.'28$/8k$\times$8k & $2.'64$/2k$\times$2k \\
Pixel size/Nyquist $\lambda$ 	& $0.''0386$/3 $\mu$m & $0.''0772$/6 $\mu$m  \\
 Wave front error/Strehl		& 150 nm rms/0.8& 150 nm rms/0.9 \\
Throughput		& $> 0.7$	& $> 0.6$ \\	
Sensitivity$^a$ for $R^b=1/5/100$ & 0.2/1/35 nJy & 13/65/1300 nJy \\
\hline
\end{tabular}

$^a$ SNR = 10 for a $10^5$~s 
integration. All spectral channels are
obtained simultaneously. \\ 
$^b$ $R$ is the number of simultaneous
spectral channels in the band-pass.
\label{capabilities}
\end{table}
\end{center}
}

\section{Unique Attributes of an IFTS}

The signal-to-noise (SNR) performance of 3-d imaging spectrometers
equipped with 2-d FPAs is the same for all architectures, in the ideal
case of photon shot noise limted operation (Bennett 1995, 2000). This
is correct so long as the spectrometers are equipped with the same
size FPAs, and the same spatial and spectral degrees of freedom of
the astronomical scene are observed.

An IFTS measures the spectrum of every pixel on the field of view and
hence provides the only efficient means of conducting {\it unbiased}
spectroscopic surveys of the high-$z$ universe, i.e., without object
preselection (e.g., using broad band colors) and without the
restrictions imposed by spectrometer slit geometry and placement.
Because there is no slit, an IFTS can record spectra for adjacent
individual objects, down to the confusion limit (e.g., resolved stellar
populations), and for moving targets (e.g., Kuiper Belt objects).

\begin{figure}[b]
\plottwo{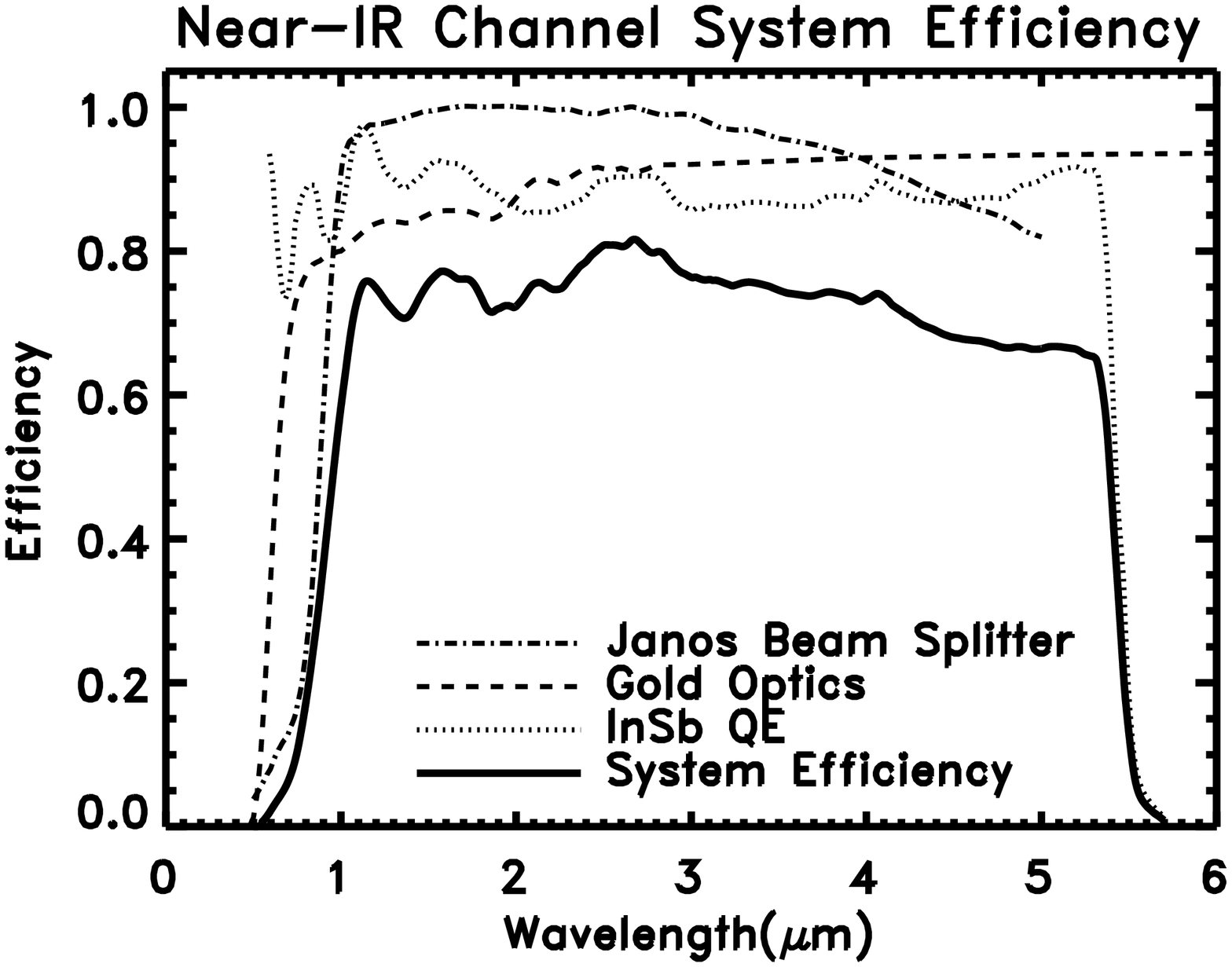}{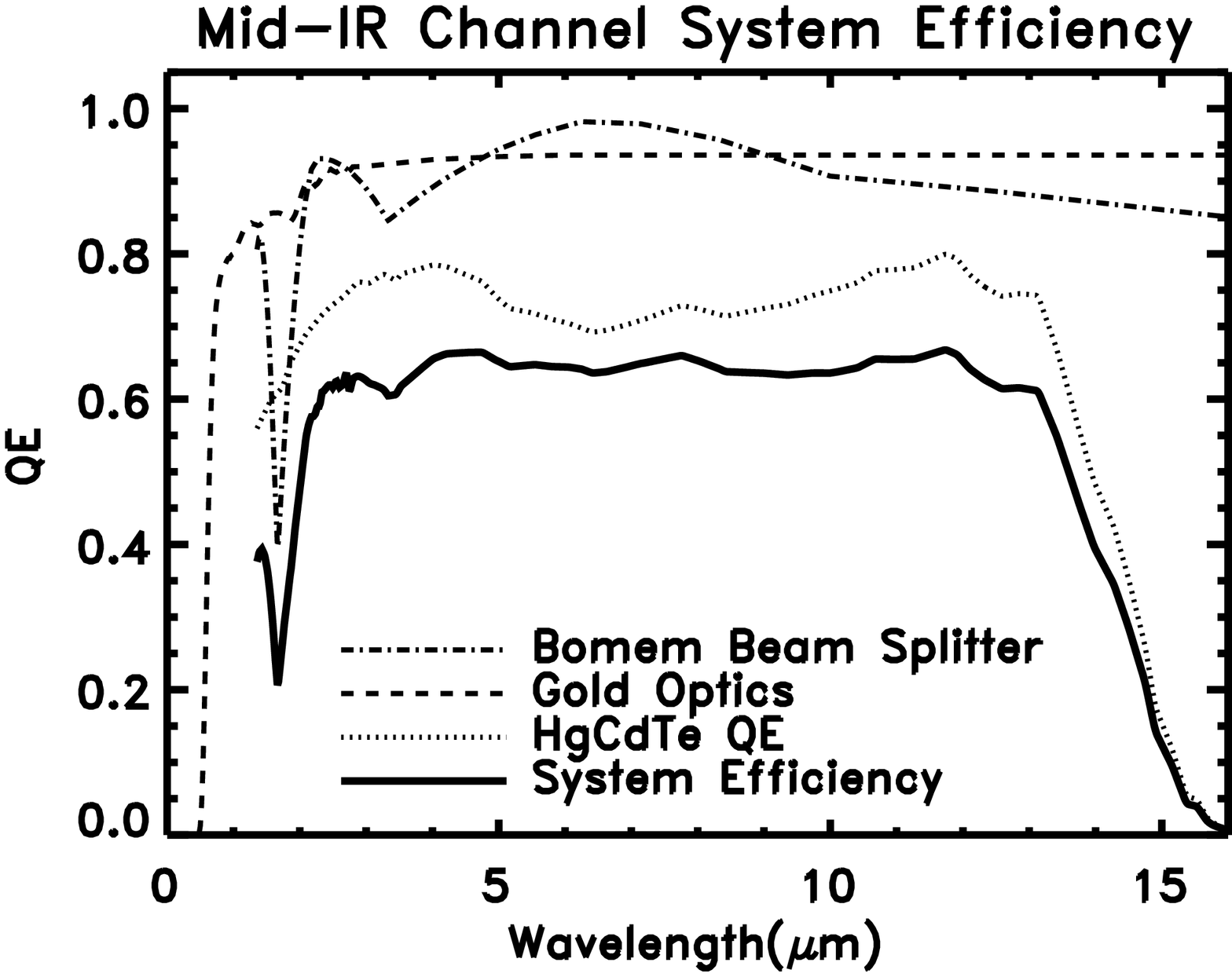}
\caption{The throughput of IFIRS for the near- and mid-IR
channels. }
\label{system.efficiency}
\end{figure}

An IFTS also allows spectroscopy over a wide bandpass, typically three
octaves, and affords flexibility in choice of resolution ($R$ =
1-$10^4$).  IFIRS uses all reflecting optics, and high-efficiency
beam-splitters.  The throughput is high and flat across the band-pass.
For the near-IR channel the efficiency is approximately 75\%,
including optics, beam splitter, and detector QE (see Fig
\ref{system.efficiency}). 

The IFTS has the brightest imagery of any imaging spectrometer design
because the full band-pass is transmitted to the FPA.  The summed
signal from the two output ports is unmodulated and corresponds to the
total broad-band photon flux entering the instrument. This
pan-chromatic image is peculiar to a four port interferometer. Thus
deep, broad-band imaging is acquired simultaneously with higher
spectral resolution data over a broad wavelength range.  A
pan-chromatic image can be formed with a filter-wheel camera by
summing the sequence of filter images, but the IFTS pan-chromatic
image has a speed advantage factor that is equal to the number of
filters used.  The IFTS advantages due to broad-band operation are
high SNR pan-chromatic imaging for science and also for telescope
guiding, tolerance of cosmic rays hits, detector noise, and internal
instrument background, and high SNR determination of flat-fields and
detector non-linearity.

A four port Michelson interferometer is intrinsically a superb
instrument from a calibration point of view.  The measured
interferogram results from the difference between spectra of sources
at the two input ports.  Calibration sources placed at the second
input port act as transfer standards for full radiometric calibrations
performed on the ground prior to flight.  The mid-IR channel is
calibrated by varying the temperature of a cold blackbody at the
second input that fills the field of view.  The near-IR channel is
calibrated using a dilute, hot blackbody, that does not produce
excessive heating of the instrument. In both cases, a full calibration
of each FPA pixel's offset, spectral responsivity, and non-linearity
can be conducted by varying integration times or source intensities.
This capability is invaluable should the system response change due to
exposure to the space environment.

IFIRS also has a hybrid, or dispersed FTS mode. In a regular FTS,
spectral information is encoded in the z-direction of the data cube,
and there is no mixing of spectral and spatial information. The
advantage is that a spectrum is recorded for every pixel on the
sky. The penalty is that the photon shot-noise from all spectral
channels is present at each frequency. This shot-noise can be reduced
by masking the telescope focal plane around objects of interest with a
programmable focal plane mask and inserting a prism into the
collimated space.  The dispersed FTS mode is used to obtain the
highest possible sensitivity at high spectral resolution
($R$=600-10,000) (Bennett, 2000).  The slit width does not determine
the spectral resolution in the dispersed FTS mode, since the spectral
resolution is derived from the interferograms. The dispersed FTS
data-cube contains spectra which are tilted with respect to the
z-axis. The tilt angle is the arc tangent of the ratio of spectral
resolutions of the dispersive element and interferometer.  The
dispersed FTS has better SNR performance than the pure FTS, and the
source density of object slits is higher than for the pure
multi-object spectrometer.

\begin{figure}[t]
\plotone{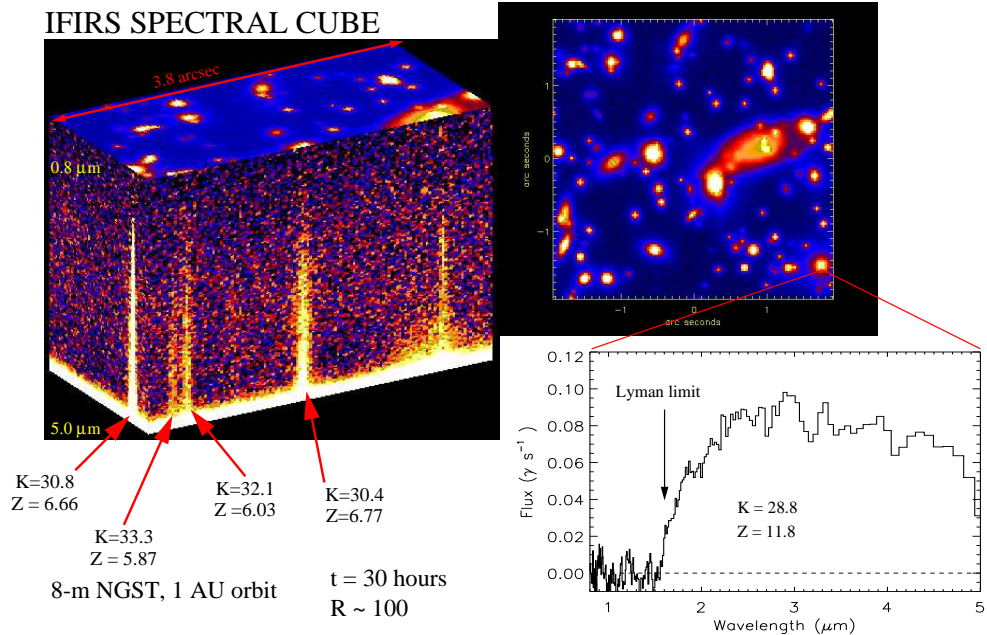}
\caption{IFIRS data cube for the Im \& Stockman (1998)
NGST deep field.}
\label{im}
\end{figure}

\section{Performance of IFIRS on NGST}

IFIRS records a spectrum for every pixel in the field of view.  This
means that data produced by IFIRS is three-dimensional, i.e., it forms
a data cube consisting of two spatial dimensions $(x,y)$ and one
wavelength dimension $(z)$.  A horizontal $(x,y)$ slice through a data
cube corresponds to a monochromatic image.  A line extracted from the
data cube in the $z$-direction corresponds to a spectrum.

To estimate the performance of an IFTS in the NGST environment we have
simulated data as follow.  An astronomical scene is represented as a
noise-free distribution of objects. This input data cube is convolved
with the telescope point spread function (PSF). This is a 3-d
convolution, since the PSF depends on wavelength. The spectra are then
multiplied by the wavelength dependent throughput.  From this spectral
data cube the interferogram is calculated by a Fourier transform. At
each OPD step noise is added. The noise sources treated are photon
shot noise from the zodiacal light, photon shot noise due to thermal
emission from the telescope, photon shot noise from the target, shot
noise due to detector dark current, and detector read-noise.  The
noisy interferogram cube is then Fourier transformed back into a
spectral data cube.

We have used Im \& Stockman's (1998) simulation of a blank field at
very low flux levels to demonstrate the spatial multiplex advantage of
IFIRS. In this simulation we show only 0.02\% of the field of view of
IFIRS. If we had simulated the entire IFIRS data cube it would contain
spectra for tens of thousands of galaxies.

The spectral energy distributions omit nebular emission lines, so
redshifts must be deduced from stellar absorption features and
spectral breaks. Simulations of star forming galaxies show that a
cluster forming stars at 2.5 $M_\odot$ yr$^{-1}$ for $10^7$ yr can be
detected in Ly$\alpha$ emission to $z=12$.  In this example objects
with K = 31.3 AB mags are detected with $SNR = 5$ per spectral
resolution element, which is sufficient SNR to identify the Lyman
break in objects as distant as z = 12.

I am indebted to the IFIRS team for their contributions to this
report: M. Abrams, C. Bennett, J. Carr, K. Cook, A. Dey, R. Hertel,
N. Macoy, S. Morris, J. Najita, A. Villemaire, E. Wishnow, \&
R. Wurtz.  This work was support by NASA.

\end{document}